\documentclass[aps,twocolumn,showpacs,showkeys,nofootinbib]{revtex4-1}

\usepackage{amsmath}
\usepackage{amsfonts,amssymb}
\usepackage{multirow}
\usepackage{graphicx,booktabs,rotating}
\usepackage{float}
\usepackage{appendix}
\usepackage{color}
\usepackage{url}
\usepackage{subfigure}
\usepackage{footnote}
\usepackage{ mathrsfs }

\def\beq{\begin{equation}}
\def\eeq{\end{equation}}
\def\bea{\begin{eqnarray}}
\def\eea{\end{eqnarray}}

\def\nl{\nonumber\\}

\def\chic1{\chi_{c1}}

\def\ppbar{p \bar p}

\def \Ds {D_s^+}

\newcommand{\gsim}{\lower.7ex\hbox{$
\;\stackrel{\textstyle>}{\sim}\;$}}
\newcommand{\lsim}{\lower.7ex\hbox{$
\;\stackrel{\textstyle<}{\sim}\;$}}

\newcommand{\eod}{\end{document}}

\large

\RequirePackage{xspace} \allowdisplaybreaks

\begin{document}

\title{The semileptonic baryonic decay $D_s^+\to p\bar p e^+ \nu_e$}
\author{Hai-Yang Cheng}
\author{Xian-Wei Kang}
\affiliation{Institute of Physics, Academia Sinica, Taipei, Taiwan
115}

\begin{abstract}
The decay $D_s^+\to p \bar p e^+\nu_e$  with a proton-antiproton
pair in the final state is unique in the sense that it is the only
semileptonic baryonic decay which is physically allowed in the
charmed meson sector. Its measurement will test our basic knowledge
on semileptonic $D_s^+$ decays and the low-energy $p\bar p$
interactions. Taking into account the major intermediate state
contributions from $\eta, \eta', f_0(980)$ and $X(1835)$, we find
that its branching fraction is at the level of $10^{-9} \sim
10^{-8}$. The location and the nature of $X(1835)$ state are crucial
for the precise determination of the branching fraction. We wish to
trigger a new round of a careful study with the upcoming more data
in BESIII as well as the future super tau-charm factory.

\end{abstract}¡¡

\maketitle

\section{Introduction}
\label{Sec:Introduction}
A great deal of effort has been devoted to the baryonic decay modes of $B$ mesons \cite{PDG,Cheng2006} due to the fact that the $B$ meson is heavy enough to allow a baryon-antibaryon pair production in the final state.
Concerning the semileptonic decay involving a baryon-antibaryon
pair, $B^-\to p\bar p \ell^- \bar\nu_\ell$ ($\ell=e,\mu)$ is  the
only measurement that has been done by the Belle Collaboration in
2014 \cite{Belle2014Bppbarlnu}. Its branching fraction was reported
to be $(5.8^{+2.4}_{-2.1}\pm0.9)\times10^{-6}$ with the upper limit
$9.6\times 10^{-6}$ at the 90\% confidence
level. In the charmed meson sector, $D_s^+\to p\bar n$ is the only
hadronic baryonic $D$ decay mode which is physically allowed. Its
branching ratio is naively expected to be very small, of order
$10^{-6}$, due to chiral suppression \cite{Pham}. Hence, the
observation of this mode by CLEO with $\mathcal{B}(D_s^+ \to p \bar
n)=(1.30\pm0.36^{+0.12}_{-0.16})\times 10^{-3}$ \cite{CLEODspnbar}
is indeed a surprise. Nevertheless, it can be explained by the
final-state rescattering of $\pi^+\eta^{(')}$ and $K^+\bar K^0$ into
$p\bar n$ \cite{ChengDspnbar}. Besides the channel $\Ds\to p\bar n$,
we notice that there is another physically allowed one, $\Ds\to
p\bar p e^+\nu_e$. The mass difference $m_{\Ds}-2m_p \approx 82$ MeV
prohibits the emission of $\pi^+$ or even the lepton $\mu^+$, thus
only the electron mode is permissible. Moreover, the $p\bar p$ pair
stays in the near-threshold region, i.e, the invariant mass squared
$s=(p_p+p_{\bar p})^2$ is not far from $4m_p^2$. The future
experimental measurement can rectify the description of $\Ds \to
\ppbar$ hadronic transition form factors as well as the low-energy
$\ppbar$ interaction. If this channel can be observed, it renders a
preponderant possibility to access the $p\bar p$ bound state due to
the low-energy $p\bar p$ region.

Below we will calculate the branching fraction of the decay channel
$\Ds\to \ppbar e^+ \nu_e$. We first consider the Cabibbo-favored
decay $\Ds \to {\bf M} (s\bar s) + e^+\nu_e$ with {\bf M} being the
meson containing a sizable $s\bar s$ quark component. Since such
meson decaying to a $\ppbar$ pair is an OZI suppressed process, we
shall focus on the intermediate mesons $\bf M$ with comparable
amount of $q\bar q=\frac{1}{\sqrt{2}}(u\bar u+d\bar d)$ and $s\bar
s$ components in order to alleviate the OZI suppression. Combining
the existing knowledge on the $\Ds\to {\bf M}$ transition form
factors and ${\bf M}\ppbar$ couplings fixed by the $p\bar p$
scattering data, we are able to take into account the $\eta, \eta',
f_0(980)$ and $X(1835)$ meson exchanges, and find that the branching
fraction of $\Ds\to \ppbar e^+\nu_e$ is at the level of $10^{-9}
\sim 10^{-8}$.

\section{Kinematics and decay rate}
The four-body decay kinematics can be described in terms of five variables:
the invariant mass squared of the $\ppbar$ pair, $s=(p_p+p_{\bar
p})^2=M_{p\bar p}^2$, the invariant mass squared of the dilepton
pair, $s_\ell=(p_\ell+p_\nu)^2$, the angles $\theta_p$,
$\theta_\ell$ and $\phi$, where $\theta_p$ ($\theta_\ell$) is formed
by the proton $p$ ($e^+$) direction in the diproton (dilepton)
center-of-mass (CMS) frame with respect to the diproton (dilepton)
line of flight in the $\Ds$ frame, and $\phi$ is the dihedral angle
between the diproton and dilepton planes. Their physical ranges are
\begin{eqnarray}
4m_p^2 &\leq& s \leq (m_{\Ds}-m_l)^2~,\nl m_l^2 &\leq& s_l\leq
(m_{\Ds}-\sqrt{s})^2~,\nl 0&\leq& \theta_p,\,\theta_l\leq \pi,\quad
0\leq \phi\leq 2\pi~.
\end{eqnarray}
One may refer to e.g., Ref.~\cite{KangBl4} for an illustration of
the four-body decay kinematics.

Instead of the separate momenta $p_p,\,p_{\bar p},\,
p_e,\,p_\nu$, it is more convenient to use the following kinematic variables
\begin{eqnarray}\label{PQLN}
P&=&p_p+p_{\bar p},\quad Q=p_p-p_{\bar p}~,\nl L&=&p_e+p_{\nu},\quad
N=p_e-p_{\nu}.
\end{eqnarray}
It follows that
\begin{eqnarray}
P^2&=&s, \,\, Q^2=4m_p^2-s,\,\,L^2=-N^2=s_l,\nl P\cdot
L&=&\frac{1}{2}(m_{\Ds}^2-s-s_l)~,\, P\cdot N=X\cos\theta_l,
\end{eqnarray}
where the function $X$ is defined by
\begin{eqnarray}
&&X(s, s_l)=((P\cdot L)^2-s\,
s_l)^{1/2}=\frac{1}{2}\lambda^{1/2}(m_{\Ds}^2,s,s_l)~,\nl
&&\lambda(x, y, z)=(x-y-z)^2-4y z.
\end{eqnarray}
with $m_p$ being the proton mass. The term $P\cdot N$ can be derived
by expressing the four momenta of $p,\,\bar p,\, e^+, \nu_e$ in the
rest frame of $D_s^+$ via the Lorentz transformation, see
e.g.,~\cite{Kangthesis}. Note that we have neglected the electron
mass over most of the available phase space (although $p\bar p$ sits
in the low energy region), i.e., $m_e^2/s_l \ll 1$. This has also been
checked numerically \footnote{As a cross check, we may first keep
the electron mass and retain the factor of $z_l=m_e^2/s_l$. Letting
$z_l\to 0$, we then recover Eq.~\eqref{eq:|T|^2} below. The
numerical results remain stable irrespective of the tiny electron
mass.}.

The decay amplitude of $D_s^+\to \ppbar e^+\nu_e$ can be written as
\begin{eqnarray}
T&=&\frac{G_F}{\sqrt{2}} V_{cs} l_\mu h^\mu~,\nl l_\mu&=&\bar
u(p_\nu) \gamma_\mu(1-\gamma^5)v(p_e)~,\nl h^\mu&=&\langle \ppbar |
V^\mu-A^\mu | D_s^+\rangle,
\end{eqnarray}
where the currents $V^\mu$ and
$A^\mu$ denote the vector and axial-vector ones, respectively, and
their hadronic matrix elements will be discussed in
Sec.~\ref{Sec:Hadronic}. We then have the differential decay rate
\begin{eqnarray} \label{eq:4PS}
d^5\Gamma &=& \frac{1}{4(4\pi)^6 m_{\Ds}^3}\sigma(s) X(s, s_l)
\sum_{\text{spins}}|T|^2 \nl &&\quad \times ds\, ds_l\,
d\cos\theta_p\, d\cos\theta_l\, d\phi,
\end{eqnarray}
with
\begin{eqnarray}
\sigma(s)=\sqrt{1-4m_p^2/s}~.
\end{eqnarray}
The four-body phase space was studied very early in 1960s within the
context of $K_{l4}$ analysis \cite{Cabibbo}, see also
Ref.~\cite{BijnensKl4} for a modern compilation. More details of derivation
can be found in e.g., Refs.~\cite{ChengYan, Kang3872}.
Equation \eqref{eq:4PS} is in agreement with
Refs.~\cite{WiseDl4,GengCQ}, as has been checked.

\section{Hadronic matrix elements and Results}
\label{Sec:Hadronic} 
\begin{figure}[h]
\centering
\includegraphics[height=78mm]{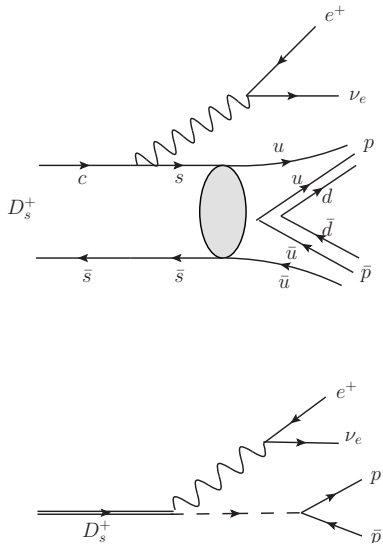}
\caption{Diagrams for the $D_s^+\to \ppbar e^+\nu_e$ decay via meson
exchanges. The upper panel depicts the quark diagram with a bulk
denoting the neutral meson with $q\bar q$ and $s\bar s$ components,
while the lower one is the meson exchange (denoted by dashed line)
diagram at the hadron level.} \label{Fig:mechanism1}
\end{figure}

We begin with the hadronic
matrix elements \cite{ChenCheng}
\begin{eqnarray} \label{eq:FFs}
\langle \ppbar | V^\mu |D_s^+\rangle &=&-i \bar u(p_p)[g_1\gamma^\mu
+ i g_2 \sigma_{\mu\nu}L^\nu + g_3 L^\mu\nl &&\qquad\quad + g_4
P^\mu +g_5 Q^\mu]\gamma^5 v(p_{\bar p})~,\nl \langle \ppbar | A^\mu
|D_s^+\rangle &=&-i \bar u(p_p)[f_1\gamma^\mu + i f_2
\sigma_{\mu\nu}L^\nu + f_3 L^\mu\nl &&\qquad\quad + f_4 P^\mu +f_5
Q^\mu] v(p_{\bar p})~.
\end{eqnarray}
Note the spinors $u$ and $v$ have a relative opposite sign under
parity transformation. Various form factors $f_i$ and $g_i$ will be
evaluated below. As mentioned in the Introduction, to alleviate the
OZI suppression for the intermediate meson exchange that leads to
the decay $D_s^+\to \ppbar e^+\nu_e$, we shall focus on the
intermediate states which have comparable $q\bar q$ and $s\bar s$
components. The two- and multi-meson exchanges are expected to be
loop suppressed, and also the direct $D_s^+p\bar p W$ production vertex without any meson exchange can be safely neglected.  We then concentrate on one-meson exchange denoted by
${\bf M}$. The combination of the existing knowledge of $D_s^+ \to {\bf M}$
transition and the coupling $\ppbar {\bf M}$ constitutes our basic
strategy. In Ref.~\cite{BRD}, we have explored the form factors and
branching fractions for the semileptonic $D_s\to {\bf M}$
transition. As for the $\ppbar {\bf M}$ part, we shall stick to the
J\"ulich nucleon-antinucleon model \cite{ppbar} \footnote{The
$\ppbar$ interaction within the framework of chiral effective field
theory involving pion degrees of freedom and contact terms was
recently explored in Ref.~\cite{Kangppbar} and Ref.~\cite{DaiLY},
see also a short review \cite{Kangrev}. A similar method has been
recently applied to charmed baryon scattering \cite{Dai4630}.} which
provides a fair description of $\ppbar$ total, elastic,
charge-exchange and annihilation cross sections. In such a $p\bar p$
model, the exchanged mesons with mass up to 1.5 GeV were considered.
We first include the spin-0 boson, $\eta,\,\eta',\,f_0(980)$ in our
study. The decay mechanism is shown in Fig.~\ref{Fig:mechanism1},
where the upper panel describes the mechanism at the quark level with
the bulk denoting the meson with the $q\bar q$ and $s\bar s$
components, and the lower one from the viewpoint of effective meson
theory with the dashed line denoting the exchanged mesons.

We will calculate the Feynman diagram to single out the contributions
of $\eta,\,\eta',\,f_0(980)$ to the $D_s^+\to \ppbar$ transition form factors.
We have
\begin{eqnarray}
\langle \ppbar |V^\mu-A^\mu |D_s^+\rangle =\sum_{\bf M} \langle {\bf
M} |V^\mu-A^\mu |D_s^+\rangle \frac{i}{p^2-m^2} \mathcal{V}, \nl
\end{eqnarray}
which amounts to inserting the intermediate meson with the momentum
$p$ and mass $m$, and $\mathcal{V}$ is the vertex of $\ppbar {\bf
M}$ coupling. The $f_0(980)$ has a large width which may remind us
of replacing $p^2-m^2$ by  $p^2-m^2-im_{f_0}\Gamma_{f_0}$. However,
it is not necessary to do so since the mass of $f_0(980)$ is still
far from the $\ppbar$ invariant mass. The induced difference by
including $\Gamma_{f_0}$ is only of order  0.1\%. The $D_s^+\to {\bf
M}$ transition \footnote{The semileptonic $D_s\to\eta$ decay can be
also treated in $SU(3)$ heavy meson chiral perturbation theory
\cite{Yan,Wise,Donoghue}, but the expression of form factors there
is valid only in the soft $\eta$ region. The pole model employed in
the current work is applicable to the whole phase space.} can be
described by
\begin{eqnarray}
\langle P|V^\mu|D_s^+\rangle
&=&\left(p^\mu-\frac{m_{\Ds}^2-m_P^2}{q^2}q^\mu\right)F_1^{D_s\to
P}(q^2)\nl &&+\frac{m_{\Ds}^2-m_P^2}{q^2}q^\mu F_0^{D_s\to P}(q^2)~,\\
\langle S |A^\mu|D_s^+\rangle
&=&-i\bigg[\left(p^\mu-\frac{m_{\Ds}^2-m_P^2}{q^2}q^\mu\right)F_1^{D_s\to
S}(q^2)\nl &&+\frac{m_{\Ds}^2-m_P^2}{q^2}q^\mu F_0^{D_s\to
S}(q^2)\bigg]~,
\end{eqnarray}
where  $P$ denotes the pseudoscalars $\eta$ and $\eta'$, $S$ the
scalar $f_0(980)$, $p=p_{\Ds}+p_p+p_{\bar p}$ and
$q=p_{\Ds}-p_p-p_{\bar p}=L$, thus $q^2=s_l$. The form factors
$F_1(q^2)$ and $F_0(q^2)$ for $D_s^+\to \eta (\eta')$ have been
investigated using the covariant light-front quark model
\cite{Chua,Verma},
\begin{eqnarray}
F_1^{D_s\to
\eta_s}(q^2)&=&\frac{0.76}{1-1.02\frac{q^2}{m_{D_s}^2}+0.40\left(\frac{q^2}{m^2_{D_s}}\right)^2}~,\nl
F_0^{D_s\to
\eta_s}(q^2)&=&\frac{0.76}{1-0.60\frac{q^2}{m_{D_s}^2}+0.04\left(\frac{q^2}{m^2_{D_s}}\right)^2}~,\nl
F_i^{D_s\to \eta}(q^2)&=&-\sin\phi F_i^{D_s\to \eta_s}(q^2)~, \nl
F_i^{D_s\to \eta'}(q^2)&=&\cos\phi F_i^{D_s\to \eta_s}(q^2)~,
\end{eqnarray}
for $i=0$ or 1. $\phi$ is the mixing angle between $\eta$ and
$\eta'$ defined by \cite{Kroll}
\begin{eqnarray}
|\eta\rangle &=& \cos\phi|\eta_q\rangle-\sin\phi|\eta_s\rangle, \nl
|\eta'\rangle &=& \sin\phi|\eta_q\rangle+\cos\phi|\eta_s\rangle.
\end{eqnarray}
It is determined to be $39.3^\circ\pm 1.0^\circ$ in the
Feldmann-Kroll-Stech mixing scheme \cite{Kroll}, which is consistent
with the recent result $\phi=42^\circ\pm2.8^\circ$ extracted from
the CLEO data \cite{Hietala}. In Ref.~\cite{BRD} it has been shown
that such a description of form factors gives a rather good
description of the branching fraction compared to experiment, and
that replacing $m_{D_s}$ by $m_D$ in the denominator does not make
significant difference for the result. As we have already commented
in Ref.~\cite{BRD}, the $D_s^+\to f_0(980)$ transition form factor
cannot be appropriately treated by the covariant light-front model
since i) $f_0(980)$ is widely believed to be a tetraquark state (see
e.g., \cite{Achasov}) or a $K\bar K$ molecular  (see e.g.
Refs.~\cite{DaiLY1706,DaiLY2gamma}) rather than a pure
quark-antiquark meson; and ii) the decay constant of $f_0(980)$
vanishes due to the charge conjugation invariance and thus there is
no reliable constraint on the parameter in its wave function within the
light-front quark model. However, the information of $F_1(q^2)$ for
$D_s^+\to f_0(980)$ is directly accessible by experiment, that is
\footnote{The value $F_1(0)=0.4$ is not shown explicitly in
Ref.~\cite{CLEO}, but can be obtained using the masses $m_{f_0}$,
$M_{\text{pole}}$ and $\mathcal{B}(D_s^+\to f_0(980)e^+\nu_e)\approx
0.4\% $ reported there. The slope of $F_1(q^2)$, namely,
$M_{\text{pole}}$, is fitted to the measured event distribution,
which differs from the $d\Gamma/dq^2$ only by an overall constant,
so $F_1(0)$ cannot be constrained by the event distribution and is left
as a float in Ref.~\cite{CLEO}.}
\begin{eqnarray}
F_1^{D_s\to f_0(980)}(q^2)=\frac{0.4}{1-q^2/M_{\text{pole}}^2}
\end{eqnarray}
with $M_{\text{pole}}=1.7^{+4.5}_{-0.7}$ GeV from the CLEO
Collaboration \cite{CLEO}. This situation is different from
Refs.~\cite{Cheng2003,Cheng2010}, where only the form factor $F_0(q^2)$
enters in the factorization scheme
of the two-body nonleptonic decay.  In the decay rate of the semileptonic decay for $D$ or
$D_s^+$ to spin-0 boson, the form factor $F_0(q^2)$ is accompanied
by the electron mass and thus negligible. In other words,
$F_0(q^2)$ can be constrained by the corresponding nonleptonic decay
rate based on factorization, but not from a direct experimental
measurement. That is \cite{Cheng2003},
\begin{eqnarray}
F_0^{D_s\to f_0(980)}(q^2)=\frac{0.52}{1-q^2/m_{D_{s1}(2536)}^2}~.
\end{eqnarray}

For the part of the $\ppbar$ interaction, we have the Lagrangian
\cite{ppbar}
\begin{eqnarray}
\mathscr L_P=  g_{P} \bar\psi(x) i\gamma^5\psi(x) \phi(x)
\end{eqnarray}
for the nucleon-nucleon-pseudoscalar ($N\!N\!P$) coupling, and
\begin{eqnarray}
\mathscr L_S=  g_{S} \bar\psi(x) \psi(x) \phi(x)
\end{eqnarray}
for the nucleon-nucleon-scalar coupling ($N\!N\!S$), with $\psi(x)$
denoting the nucleon field and $\phi(x)$ the meson field. The
dimensionless couplings read \cite{ppbar} \footnote{Note that
the factor $4\pi$ is sometimes absorbed in the couplings.}
\begin{eqnarray}
g_{\eta}=2.87,\quad g_{\eta'}=3.72,\quad g_{f_0}=8.48.
\end{eqnarray}
Note for the $N\!N\!P$ coupling there is another form, namely, the
so-called pseudovector coupling,
\begin{eqnarray}
\mathscr
L_{\rm{pv}}=\frac{f}{m_\pi}\bar\psi(x)\gamma^5\gamma^\mu\psi(x)\cdot
\partial_\mu\phi(x).
\end{eqnarray}
The pseudoscalar coupling and the pseudovector one are related by
\begin{eqnarray}
\frac{f}{m_\pi}=\frac{g_{P}}{2m_N}
\end{eqnarray}
for free nucleon satisfying the Dirac equation. One may refer to
Ref.~\cite{Weise} for more details. The pseudoscalar coupling was
used in Ref.~\cite{ppbar}, although the pseudovector form of the
Lagrangian appeared in the appendix of the paper.

Another essential and ``dominant'' piece should be the $X(1835)$
($J^{PC}=0^{-+}$) exchange since i) it locates near the $p\bar p$
threshold such that the propagator can enhance the contribution
\footnote{The quark-hadron duality is very subtle in the
near-threshold region \cite{Bigi2017}. Owing to the large width of
$X(1835)$, its wave function is overlapped with the $p\bar p$ state
to some extent.}, and ii) the strong connection/relation between
$X(1835)$ and the $p\bar p$ state. The first observation of
$X(1835)$ (denoted by the $X$ particle below) was reported by BESII
from the channel $J/\psi\to\gamma p\bar p$ \cite{BESII}, where the
mass reads $1859^{+3\, +5}_{-10\,-25}$ MeV with the statistic and
systematic errors, in order, by using the $S-$wave Breit-Wigner
function. The huge enhancement of the event distribution near the
$p\bar p$ threshold was interpreted as the effect due to the $p\bar
p$ final-state interaction (FSI) \cite{Sibirtsev}, where the
Watson-Migdal approach is exploited, i.e., the amplitude for
$J/\psi\to\gamma p\bar p$ is expressed by a normalization constant
multiplied by the $p\bar p$ scattering $T$-matrix. A refit with the
inclusion of the fixed FSI factor introduced in
Ref.~\cite{Sibirtsev} has been carried out in a subsequent
publication \cite{BES2005}.  The resulting mass is slightly changed and reads
$1826.5^{+13.0}_{-3.4}$ MeV \cite{PDG}. From the state-of-the-art
viewpoint, such FSI treatment has been superseded by the outcome of
Ref.~\cite{Kangppbar}, where the total amplitude $A$ is written as
$A=A_0+A_0 G_0 T$ with $A_0, \,G_0, \,T$ denoting bare production
amplitude without FSI, free Green function and $p\bar p$ scattering
$T$-matrix, respectively. One can refer to the review in
Ref.~\cite{Kangrev} for more details. It has been shown that the
threshold enhancement could be indeed a $p\bar p$ bound state
\cite{Kangppbar}. However, one should be cautious that the pure FSI
explanation proposed in Ref.~\cite{Sibirtsev} reproduces the data
very well and thus cannot be excluded. To date, the nature or even
its existence of $X(1835)$ still remains mysterious. However,
$X(1835)$ can be viewed, at least, as a poor man's approach or an
effective way to incorporate the strong FSI of $p\bar p$, and in this
respect, we include it as a subthreshold resonance in our
meson-exchange model calculation.

Note that the $X(1835)$ has also been observed in $\gamma \eta'
\pi\pi$ channel with a statistical significance of $7.7\sigma$
\cite{BES2005}. So, it could be most likely a mixing state of $p\bar
p$ and $s\bar s$ and this idea has been investigated in e.g.,
Ref.~\cite{YanML}. Then one may write
\begin{eqnarray}\label{eq:X1835mixing}
| X(1835)\rangle = c_1 |p\bar p\rangle + c_2 |s\bar s\rangle,
\end{eqnarray}
with $|c_1|^2+|c_2|^2=1$. The maximum production for $D_s^+\to p\bar
p e^+ \nu_e$ corresponds to $c_1=c_2=1/\sqrt{2}$. The Lagrangian for
the $X(1835)p\bar p$ coupling is of the same form as $p\bar p\eta$,
and we will take the coupling constant $g_{Xp\bar p}\approx 3.5$
\cite{ZhuSL} provided that $X(1835)$ is a pure baryonium. This value
agrees with the one given in Ref.~\cite{MaYL} after applying the
Weinberg compositeness theorem
\cite{Weinberg,GuoOller,KangZb,Kang3872}, i.e., the coupling
$g_{Xp\bar p}$ is obtained from the vanishing wave function
renormalization. In the case of Eq.~\eqref{eq:X1835mixing}, the true
$Xp\bar p$ coupling will be multiplied by a factor of $1/\sqrt{2}$,
while the transition $D_s^+\to X(1835)$ will proceed via the form
factors $F_1^{D_s\to \eta_s}/\sqrt{2}$ and $F_0^{D_s\to
\eta_s}/\sqrt{2}$ \cite{BRD,Chua,Verma} \footnote{Under these
discussions, we can also obtain an approximate branching fraction
$D_s^+\to X(1835) e^+\nu_e\approx(1.6^{+0.2}_{-0.7})\times 10^{-6}$
using the averaged mass $m_X=1826.5^{+13.0}_{-3.4}$ MeV obtained by
PDG \cite{PDG}.}.

Combining all these ingredients together, we obtain
\begin{widetext}
\begin{eqnarray}
g_3&=&\frac{g_\eta}{s-m_\eta^2}\Big[(1-\Delta_\eta)F_1^{D_s\to\eta}(s_l)+\Delta_\eta
F_0^{D_s\to\eta}(s_l)\Big]\nl
&+&\frac{g_{\eta'}}{s-m^2_{\eta'}}\Big[(1-\Delta_{\eta'})F_1^{D_s\to\eta'}(s_l)+\Delta_{\eta'}
F_0^{D_s\to\eta'}(s_l)\Big]\nl &+&\frac{g_{Xp\bar
p}/\sqrt{2}}{s-m_X^2}\Big[(1-\Delta_X)F_1^{D_s\to X}(s_l)+\Delta_X
F_0^{D_s\to X}(s_l)\Big]~,\\
f_3&=&\frac{g_{f_0}}{s-m^2_{f_0}}\Big[(1-\Delta_{f_0})F_1^{D_s\to
f_0}(s_l)+\Delta_{f_0}F_0^{D_s\to f_0}(s_l)\Big]~,\\
g_4&=&\frac{2g_\eta}{s-m_\eta^2}F_1^{D_s\to\eta}(s_l)+\frac{2g_{\eta'}}{s-m_{\eta'}^2}F_1^{D_s\to\eta'}(s_l)
+\frac{2g_{Xp\bar p}/\sqrt{2}}{s-m_X^2}F_1^{D_s\to X}(s_l)~,\\
f_4&=&\frac{2g_{f_0}}{s-m_{f_0}^2}F_1^{D_s\to f_0}(s_l)~.
\end{eqnarray}
\end{widetext}
and all other form factors vanish, where
\begin{eqnarray}
\Delta_{S[P]} =\frac{m_{D_s}^2-m^2_{S[P]}}{s_l}.
\end{eqnarray}
To evaluate the amplitude modulus squared, we introduce the hadronic and leptonic tensor currents given by
\begin{eqnarray}
\mathcal{H}^{\mu\nu}&=&h^\mu h^{\nu*}\nl
&=&2\Big[\left(s-4m_p^2\right)\big(f_3^*f_4P^\mu L^\nu +
f_3f_4^*P^\nu L^\mu \nl &&\qquad\qquad\qquad+|f_3|^2L^\mu L^\nu +
|f_4|^2 P^\mu P^\nu\big)\nl && +s \big(g_3^*g_4P^\mu L^\nu+g_3 g_4^*
P^\nu L^\mu\nl &&\qquad\qquad\qquad+|g_3|^2L^\mu L^\nu+|g_4|^2P^\mu
P^\nu\big)\Big]~,
\end{eqnarray}
and
\begin{eqnarray}
\mathcal{L}_{\mu\nu}&=&l_\mu l_\nu^*\nl &=&4[(L^\mu L^\nu-N^\mu
N^\nu)-g^{\mu\nu}(s_l-m_l^2)\nl &&\quad
+i\epsilon_{\mu\nu\sigma\tau}L^\sigma N^\tau],
\end{eqnarray}
respectively, with the convention $\epsilon^{0123}=1$.
The amplitude modulus squared
reads
\begin{eqnarray}\label{eq:|T|^2}
|T|^2&=&G_F^2|V_{cs}|^2\big(|f_4|^2(s-4m_p^2)+s|g_4|^2\big)\nl
&&\times\Big(m_{\Ds}^4+(s+s_l)(s-s_l-2m_{\Ds}^2)\nl&&\quad-4X^2(s,
s_l)\cos^2\theta_l\Big),
\end{eqnarray}
where the $g_3$ and $f_3$ terms are suppressed by the smallness of the electron
mass. This leads to the branching fraction
\begin{eqnarray}
\mathcal B(D_s^+\to p\bar p e^+\nu_e)=\frac{1}{\Gamma_{\Ds}}\int
d\Gamma^5=3.5\times10^{-9},
\end{eqnarray}
based on the mass of $1826.5$ MeV for the $X(1835)$ reported in
PDG \cite{PDG}. The uncertainties arise from various sources,
for example, the coupling constants and the mass of $X(1835)$. The
dominant uncertainty should be ascribed to the precision on the $X(1835)$
mass. If we use the mass 1859 MeV, which corresponds to the fit
without the primitive treatment of FSI as we have already discussed
above, the branching fraction will become
\begin{eqnarray}
\mathcal B (D_s^+\to p\bar p e^+\nu_e)=1.5\times 10^{-8}.
\end{eqnarray}
As noticed in passing, the $X(1835)$ exchange should dominate due to
its proximity to the $p\bar p$ state. The $X(1835)$ alone will
contribute to $1.42\times 10^{-8}$ for the branching fraction after
turning off the $\eta,\,\eta',\,f_0(980)$ effects. If the mass of
$X(1835)$ is closer to the $p\bar p$ threshold, the branching
fraction will be further increased. Considering the width of $X(1835)$
(around 80 MeV) \cite{Kangppbar}, the branching fraction will become
smaller by a few times. In this sense, we prefer to emphasize the
importance of the ``precision'' on the $X(1835)$ mass measurement.
By the end of 2018, around $10^{10}$ $J/\psi$ data samples are going
to be accumulated within the one-year running period \cite{Front}
and both $J/\psi\to \gamma p\bar p$ and $J/\psi\to
\gamma\eta'\pi\pi$ can be re-examined to improve the accuracy. The
experimental situation will be further improved in the case of super
tau-charm factory \cite{super1,super2,super3} with the planned
luminosity of 100 times as much as BESIII.

\begin{figure}[h]
\centering
\includegraphics[height=40mm]{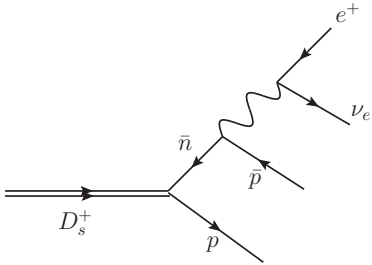}
\caption{The diagram  for baryon exhange: $D_s^+\to p \bar n$
followed by the neutron beta decay, $\bar n\to \bar p e^+ \nu_e$. It
involves two weak vertices and thus the contribution is negligible.}
\label{Fig:mechanism2}
\end{figure}

In principle, the diagram with one-baryon exchange can be also
considered. There is the process $D_s^+\to p\bar n$ followed by the
neutron beta decay, $\bar n\to \bar p e^+\nu_e$ as depicted in
Fig.~\ref{Fig:mechanism2}. As noticed before, the decay rate of
$D_s^+\to p\bar n$ measured to be $(1.3\pm0.4)\times 10^{-3}$
\cite{CLEODspnbar} is unexpectedly large beyond the naive weak
annihilation mechanism \cite{Stech,ChengDspnbar}. Moreover, both the
antineutron and antiproton in Fig.~\ref{Fig:mechanism2} are soft and
hence the contribution of this diagram may be possibly large due to
the propagator of the antineutron. However, the net contribution is
again highly suppressed because it involves two weak vertices
proportional to $G_F^2$. We indeed calculated
Fig.~\ref{Fig:mechanism2}, and found that it contributes to the form
factors $g_1,\,g_2,\,g_3$ and $f_1,\,f_2,\,f_3$. Within such
microscopic process, this gives a picture of how the general form
factors (constructed from Lorentz structure) emerge.

Here we comment on the possible OZI violation. The question may
arise from the large $\phi$ production rate in $p\bar p$ collisions
compared to the $\omega$ one, which is attributed to either the
intrinsic $s\bar s$ component in the wave function of the proton
\cite{Ellis} or the rescattering of kaons \cite{Locher,Klempt}, see
also the reviews in Refs.~\cite{phiproduction,KlemptReview}. The
strangeness content of the nucleon has also been revealed in several
experimental observations, e.g., the strange quark spin
polarization, $\sigma_{\pi N}$ term, magnetic moment of the proton and the
ratio of strange and non-strange quark flavor distributions.
However, the weight of the strange content is still small such that
it is not expected to make large influence on the current results.
On the other hand, the low-energy $p\bar p$ scattering data can be
fairly well reproduced without the inclusion of the $\phi$ meson
exchange, as e.g., done in Ref.~\cite{ppbar}, for which we stick to
the $p\bar p {\bf M}$ coupling.

In the $B$ meson sector, the semileptonic baryonic decay $B^-\to p\bar p\ell^-\bar \nu_\ell$ has
been studied in \cite{GengBppbar} where the form factors $f_i$ and $g_i$ with $i=1,\cdots,5$ defined
in analog to Eq. (\ref{eq:FFs}) were obtained by fitting them to the available data of $B\to \ppbar {\bf M}$
in conjunction with the pQCD counting rule for form factors. However, this pQCD argument is not applicable to
our case as the energy release in $D_s^+\to p\bar p$ transition is rather small. Moreover, we notice that the
predicted branching fraction ${\cal B}(B^-\to p\bar p\ell^-\bar \nu_\ell)=1.04\times 10^{-4}$ in \cite{GengBppbar}
is too large compared to the experimental observation of order $6\times 10^{-6}$.

\section{Conclusion}
\label{Sec:Conclusion} In this work we have discussed the unique decay
$D_s^+\to p \bar p e^+\nu_e$  with a proton-antiproton pair in the
final state. It is the only semileptonic baryonic decay which is
physically allowed in the charmed meson sector, besides the hadronic
baryonic decay $D_s^+\to p\bar n$. There is abundant physics in this
channel. Its measurement will test our knowledge on baryonic weak
decays and the low-energy $p\bar p$ interactions. Taking into
account the contributions from the intermediate states
$\eta,\,\eta',\,f_0(980)$ and $X(1835)$, we find that its branching
fraction is $10^{-9} \sim 10^{-8}$. Our prediction can be tested by
BESIII/BEPCII data and its measurement is ongoing.

\section*{Acknowledgments} One of the authors XWK wishes to thank Prof. Hai-Bo Li, a member of the BESIII Collaboration,
for pointing out this interesting topic.  He is also grateful to Dr.
J.~Haidenbauer for many invaluable discussions on the $p\bar p$
interaction and Mr. Xinxin Ma for very useful discussion on the
relevant experimental measurements. This work is supported by the
Ministry of Science and Technology of R.O.C. under Grant No.
104-2112-M-001-022.


\end{document}